\documentclass[aps,pra,epsfigure,twocolumn]{revtex4}
\usepackage{graphicx}
\usepackage{dcolumn}
\usepackage{amsthm}
\usepackage{bm}
\usepackage{epsfig}
\usepackage{amsmath}
\usepackage{amssymb}

\newcommand{\ket}[1]{\left\vert#1\right\rangle}
\newcommand{\modul}[1]{\left\vert#1\right\vert}

\newcommand{\one}{\mbox{$1 \hspace{-1.0mm}  {\bf l}$}}
\newcommand{\pro}[2]{\left\vert#1\right\rangle\left\langle#2\right\vert}

\newcommand{\bra}[1]{\left\langle#1\right\vert}

\begin{document}

\title{Entanglement of mixed macroscopic superpositions: an entangling-power study}

\author{M. Paternostro$^1$, H. Jeong$^2$ and M. S. Kim$^1$}

\affiliation{$^1$School of Mathematics and Physics, The Queen's University, Belfast, BT7 1NN, UK\\
$^2$Center for Quantum Computer Technology, Department of Physics,
University of Queensland, St Lucia, Qld 4072, Australia}
\date{\today}

\begin{abstract}
We investigate entanglement properties of a recently introduced class of macroscopic quantum superpositions in two-mode mixed states. One of the tools we use in order to infer the entanglement in this non-Gaussian class of states is the power to  entangle a qubit system. Our study reveals features which are hidden in a standard approach to entanglement investigation based on the uncertainty principle of the quadrature variables. We briefly describe the experimental setup corresponding to our theoretical scenario and a suitable modification of the protocol which makes our proposal realizable within the current experimental capabilities.  
\end{abstract}


\maketitle
%%%%%%%%%%%%%%%%%%%%%%%%%%%%%%%%%%%%%%%%%%%%%%%%%%%%%%

\section{Introduction}

Quantum entanglement is the key element in many applications of quantum information processing %%@
(QIP) ranging from quantum computation \cite{qc} 
%~\cite{nonlocal,jacob}
to
%quantum key distribution \cite{qkd}. 
communication \cite{communication}.
The studies 
%and characterization
of entanglement represent an active line of research in modern quantum physics which so far %%@
has found only partial answers. On the practical side, entanglement is the core of a new {\it %%@
paradigm} for quantum computation~\cite{oneway} and the catalyst for the performance of tasks %%@
which are impossible within the classical domain~\cite{tasks}. 

More recently, the relation between quantum entanglement and thermodynamical properties of %%@
macroscopic objects has become the center of an extensive study~\cite{lavorivlatko1}. The %%@
existence of long-range quantum correlations between the parties of a complex many-body system %%@
is allegedly the reason for the peculiar behavior of macroscopic properties of %%@
solids~\cite{lavorivlatko2}. In this context, it is intellectually stimulating and %%@
pragmatically very important to investigate and understand the existence of entanglement %%@
between macroscopic physical systems and the influences of temperature on it. Theoretically, %%@
some steps in this direction have been performed with the study of radiation pressure-induced %%@
entanglement between a movable mirror and the electromagnetic field of a %%@
cavity~\cite{sougato,camerino,lavorivlatko3}.

Very often, the quantitative analysis of even simplified models of interaction faces important %%@
practical difficulties related to the necessity of treating non-Gaussian states of %%@
continuous-variable (CV) systems.
%A state is said to be Gaussian if its characteristic function is
%a Gaussian in the phase space~\cite{BarnettRadmore}.
For non-Gaussian states, %on the other hand,
there is a lack of objective criteria to determine whether or not entanglement is present. %%@
In these years, there have been a few proposals designed to bypass this problem, including \cite{Hillery, Agarwal}. 
Unfortunately, so far a totally satisfactory answer has not been provided.
This serious limitation largely affects the extent to which an analysis of thermally entangled %%@
systems can be conducted. A way to bypass the problem is given by the formal restriction of %%@
the Hilbert space of the composite CV system to an effective discrete one~\cite{sougato2}, %%@
where necessary and sufficient conditions for the existence of bipartite entanglement can be %%@
used~\cite{pereshorodecki}. However,
even though this strategy is theoretically exploitable, its experimental realization is %%@
extremely hard. 
%its experimental studies are extremely hard.
%as it implies the effective projection of a CV state onto a bidimensional Hilbert space 
%which is in general a very difficult step. 

In this paper, we investigate the entanglement properties of a class of states which have been %%@
very recently introduced by Jeong and Ralph~\cite{jacob2} in order to provide a reasonable %%@
analogy of the Schr\"{o}dinger's cat paradox~\cite{gatto}. A class of entangled state in %%@
\cite{jacob2} can be considered as a generalization of the two-mode Schr\"odinger cat-like %%@
state \cite{wineland}, which corresponds to an entangled state between microscopic and %%@
macroscopic systems, to a thermal mixture. The {\it macroscopic} part of this cat-like state %%@
is represented by the displaced thermal state of a harmonic oscillator while the microscopic %%@
part is an atomic (or single photon) qubit. The other type of entangled state studied %%@
in~\cite{jacob2} corresponds to an entangled state between two macroscopic systems and can be %%@
generated {\it through an additional conditional measurement}. Such a state can be understood %%@
as a generalization of an entangled coherent state \cite{Sanders92}, which has been found %%@
useful for QIP \cite{mqip}, to a thermal mixture. For the second type of the macroscopic %%@
entangled state, it was found that Bell's inequality can be violated up to the maximum bound %%@
even though the thermal temperature becomes extremely large~\cite{jacob2}. On the other hand, %%@
the existence and degree of entanglement in the first type of the microscopic-macroscopic %%@
entangled state, which we shall call ``generalized cat-like state", remains a question to be %%@
answered. In other words, the first question that should be answered in this paper is ``How %%@
much entanglement can be generated by an interaction between microscopic quantum state (qubit) %%@
and macroscopic classical system (thermal state) without any additional process when the %%@
thermal temperature becomes extremely large?''

%Naively, one could expect that the thermal average would smear out any inter-mode
%entanglement in the state thus generated. However, strong non-classical behaviors
%of this generalized cat-like state is witnessed by the large negativity of the
%associated Wigner function
%quantum interference and 
%violations of Bell's inequality up to the maximum value 
%\cite{jacob2}. 

%How much entanglement is contained in this type of mixed entangled states is another %%@
%interesting question.

However, unfortunately,
the states which we are interested in are non-Gaussian mixed CV states for which 
the known entanglement criterions for CV states~\cite{simon} fail, leaving a degree of %%@
ambiguity which prevents any firm statement on the presence of quantum correlations. To bypass %%@
this problem, we describe an alternative theoretical method which is then corroborated by %%@
testing the {\it entangling power} of the generalized cat-like state~\cite{ioMEMS}. The test %%@
is based on the capability of a state to induce entanglement, by means of only bilocal %%@
interactions, between two initially uncorrelated qubits.  As local unitary operations alone %%@
cannot create entanglement, the entangling power provides a sufficient condition for the %%@
inseparability of the CV state being investigated. Therefore, another question that we %%@
naturally address in this paper is ``How can we transfer the entanglement generated by a %%@
microscopic-macroscopic interaction to the initially separable bipartite system of two %%@
non-interacting qubits?" 

Differently from the above mentioned effective-projection %%@
technique~\cite{lavorivlatko3,sougato2}, our procedure is operative as it is immediate to %%@
design the general scenario for an experimental investigation. This approach can be extended %%@
to other non-trivial class of non-Gaussian CV states, for example a set of states which can be generated from the generalized cat-like state by measuring %%@
the microscopic part of the superposition and subsequently unitarily manipulating the %%@
remaining macroscopic part. This results in the coherent superposition of thermally averaged %%@
two-mode states. Again, the application of the entangling power test shows ability to induce %%@
entanglement in a two-qubit system. Our analysis reinforces the ideas related to the possibility of macroscopic %%@
entanglement at non-zero temperature. We thus provide a new tool in the non-trivial problem of %%@
the quantitative analysis of the properties of non-Gaussian states which is flexible enough to %%@
reveal temperature-resilient macroscopic entanglement.

The remainder of this paper is organized as follows. In Section~\ref{gattogenerale}, we %%@
introduce a generalized cat-like state, which is a macroscopic non-local state. We first address the %%@
behavior of the state's variance matrix to show that the Simon's CV-state entanglement %%@
criterion fails in revealing any quantum correlation in this state. On the other hand, by %%@
changing picture and restricting the attention to bidimensional subsectors of the infinite %%@
dimensional Hilbert space of the CV subsystem, it is possible to infer the entanglement %%@
properties of this class of generalized cat-like states. This approach allows us to highlight %%@
the striking effects of temperature and displacement (in the phase space) over the %%@
entanglement of the state, thus providing an important quantitative insight into its %%@
properties. In Section~\ref{trasferisco}, by adopting the entangling power viewpoint, we show %%@
that entanglement can be reliably transferred to two independent qubits which have interacted %%@
with the state under investigation, giving us the possibility of constructing a highly %%@
entangled quantum channel of static qubits. On the other hand, this approach provides us with %%@
a way to detect entanglement in the CV state simply by looking at the state of two qubits. %%@
Section~\ref{experiment} describes exactly this experimental protocol and a modified one %%@
which, with limited modifications to the original idea, turns out to be %extremely
realistic. Finally, in Section \ref{conclusioni} we resume the central results of our analysis %%@
and complement it with a brief study the entanglement of thermally weighted %%@
entangled coherent states. We show that the entangling power is an exploitable tool in this case as %%@
well and reveals the presence of entanglement within a significant range of temperature. 

%%%%%%%%%%%%%%%%%%%%%%%%%%%%%%%%%%%%%%%%%%%%%%%%%%%%%%GATTOGENERALIZZATO%%%%%%%%%%%%%%%%%%%%%%%%@
%%%%%%%%%%%%%%%%%%%%
\section{Generalized cat-like states}
\label{gattogenerale}

Let us consider the interaction of a qubit of its logical states $\{\ket{0},\ket{1}\}_{m}$, %%@
initially prepared in the balanced linear superposition %%@
$\ket{+}_m=(1\sqrt{2})(\ket{0}+\ket{1})_{m}$, with the macroscopic part of the generalized %%@
cat-like state here represented by a physical system $M$ in an initial displaced thermal state %%@
$\rho^{th}_{M}(V,d)=\int{d}^2\alpha{P}^{th}_{M}(V,d)\ket{\alpha}_{M}\bra{\alpha}$. We have %%@
introduced the coherent state $\ket{\alpha}$ of its amplitude %%@
$\alpha=\modul{\alpha}e^{i\zeta}\in\mathbb{C}$ and the thermal probability distribution %%@
\begin{equation}
P^{th}_{M}(V,d)=\frac{2}{\pi(V-1)}e^{-\frac{2\modul{\alpha-d}^2}{V-1}},
\end{equation}
where $V$ is the variance of the distribution, related to the temperature of the thermal %%@
distribution by the relation $V=(e^{\beta\omega}+1)/(e^{\beta\omega}-1)$ with %%@
$\beta^{-1}=k_{B}T$. Here, $k_{B}$ is the Boltzmann constant, $T$ is the radiation temperature %%@
and $\omega$ its frequency. We have indicated with $d$ the displacement of the state from the %%@
origin of the phase-space. The relation between the variance and the mean photon number %%@
$\bar{n}$ of the field is $V=2\bar{n}+1$. The interaction is ruled by the cross-phase %%@
modulation Hamiltonian $\hat{H}_{K}=\chi\hat{m}^{\dag}\hat{m}\hat{M}^{\dag}\hat{M}$ ($\hbar=1$ is assumed throughout the paper). Here %%@
$\hat{m}$ and $\hat{m}^{\dag}$ ($\hat{M}$ and $\hat{M}^{\dag}$) are the annihilation and %%@
creation operators of the microscopic (macroscopic) system respectively and $\chi$ is the rate %%@
of non-linearity of the interaction~\cite{jacob2}. After an interaction time $t$, the %%@
dynamical evolution corresponding to the Hamiltonian $\hat{H}_{K}$ gives rise to the state 
\begin{equation}
\label{eq3}
\rho^{ent}_{mM}=\int{d}^{2}\alpha{P}^{th}_{M}(V,d)\ket{\psi}_{mM}\!\bra{\psi},
\end{equation}
with $\ket{\psi}_{mM}=(1/\sqrt{2})(\ket{0,\alpha}+\ket{1,\alpha{e}^{i\varphi}})$. Here, %%@
$\varphi=\chi{t}$ is the cross-phase shift induced by the model $\hat{H}_{K}$. 

\begin{figure}[t]
\psfig{figure=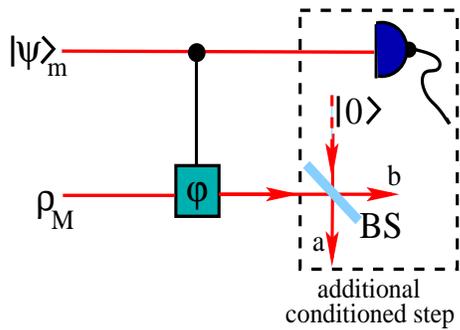,width=6.0cm,height=4.3cm}
\caption{Scheme of principle of the protocol 
%introduced in ref.~\cite{jacob2} in order
 to produce the generalized version of the cat-like state (microscopic-macroscopic entangled %%@
state) %\cite{jacob2}
  and the second class of thermally weighted entangled coherent states (performed by the %%@
additional conditioned step depicted in the dashed box). The conditional-$\varphi$ gate is the %%@
effective evolution induced by the cross-Kerr interaction between the microscopic part $m$ (in %%@
the pure state $\ket{\psi}_{m}$) and the macroscopic one in the thermal displaced state %%@
$\rho^{th}_{M}$. BS is a $50:50$ beam splitter and the symbol for a detector is also shown.}
\label{schema}
\end{figure}

In order to pursue our investigation, we find it convenient to preliminarily consider just the %%@
pure superposition $\ket{\psi}_{mM}$, whose entanglement properties we want to quantitatively %%@
address. This state has been considered as a reasonable example of Schr\"{o}dinger cat state %%@
and has been experimentally produced as an entangled state of internal and external degrees of %%@
freedom of a single trapped-ion~\cite{wineland}. Clearly, for %%@
$\langle{\alpha}\vert\alpha{e}^{i\varphi}\rangle\simeq{0}$, this state carries almost one ebit %%@
of entanglement.
However, the asymmetric non-Gaussian nature of this superposition makes the entanglement %%@
analysis of this state quite non-trivial. For instance, the entanglement does not emerge at %%@
the level of the variance matrix. The variance matrix of a two-mode CV state is defined as %%@
$V_{\mu\nu}=\langle\{\hat{x}_{\mu},\hat{x}_{\nu}\}\rangle\,(\mu,\nu=1,2)$. %%@
The variance matrix is in one-to-one correspondence with the characteristic functions of a Gaussian CV %%@
state which, in turns, gives us information about the actual state of the %%@
system~\cite{myungmunro,duan}. However, this is not true for the %%@
non-Gaussian state $\ket{\psi}_{mM}$, whose variance matrix reads $\gamma^{ent}_{mM}=
\begin{pmatrix}
{\bm A}&\bm{C}\\
\bm{C}^{T}&\bm{B}
\end{pmatrix}$
with ${\bm A}=2\one$,
\begin{equation}
\bm{B}=2\modul{\alpha}^2
\begin{pmatrix}
1+\cos\varphi\cos(2s)&\sin(2s)\cos{\varphi}\\
\sin(2s)\cos{\varphi}&1-\cos\varphi\cos(2s)
\end{pmatrix}
+\one
\end{equation}
and
\begin{equation}
\bm{C}=\alpha_{\varphi}
\begin{pmatrix}
\cos(s)\cos(r)&\sin(s)\cos(r)\\
\cos(s)\sin(r)&\sin(s)\cos(r)
\end{pmatrix}.
\end{equation}
Here we have defined $\alpha_{\varphi}=2\modul{\alpha}e^{-\modul{\alpha}^2(1-\cos\varphi)}$, %%@
$s=\zeta+\varphi/2$ and $r=\modul{\alpha}^2\sin{\varphi}+\varphi/2$. We have found that %%@
Simon's separability criterion~\cite{simon,derek} which 
is the most successful criterion for CV entanglement, is not able to show quantum correlations %%@
in this state. Interestingly enough, the criterion fails for $\varphi=\pi$ which is the value %%@
at which one would expect the largest degree of entanglement to be found between the %%@
subsystems. We have also checked that the conditions for inseparability 
established in~\cite{Hillery} is not verified by the state we are considering so that other ways have to be researched.
The above intuitive expectation about the entanglement at $\varphi=\pi$ is confirmed by the following simple analysis.
The state $\ket{\psi}_{mM}$ can be written in its Schmidt decomposition as 
\begin{equation}
\label{schmidt}
\ket{\psi}_{mM}=\sqrt{\lambda_{-}}\ket{\Phi_{+}}_m\ket{\Psi_+}_M+
\sqrt{\lambda_{+}}\ket{\Phi_{-}}_m\ket{\Psi_-}_M,
\end{equation}
where %%@
$\ket{\Phi_\pm}_m=(1/\sqrt{2})(\ket{0}\pm{e}^{2i\modul{\alpha}^2\sin{\varphi}}\ket{1})_m$
and $\ket{\Phi_\pm}_M=N_{\pm}(\ket{\alpha}\pm{e}^{-i\modul{\alpha}^2\sin{\varphi}}
\ket{\alpha{e}^{i\varphi}})_M$ with the normalization factors
\begin{equation}
N_\pm=\frac{1}{\sqrt{2(1\pm\exp[-\modul{\alpha}^2(1-\cos\varphi)])}}.
\end{equation}
The coefficients of the superposition (\ref{schmidt}) are defined in terms of the $N_{\pm}$ %%@
factors as $\lambda_{\pm}=N^{2}_{\pm}/(N^{2}_{+}+N^{2}_{-})$. In order to quantify the %%@
entanglement within a state of its density matrix $\bm{\rho}_{ab}$, we use the {\it negativity %%@
of partial transposition} (NPT)~\cite{npt}. NPT is a necessary and sufficient condition for %%@
entanglement of any bipartite qubit state~\cite{npt}. The corresponding entanglement measure %%@
is defined as ${\cal E}=\max\{0,-2\epsilon^{-}\}$ with $\epsilon^{-}$ the negative eigenvalue %%@
of the partial transposition of $\bm{\rho}_{ab}$ with respect to one of its parties.
The entanglement in $\ket{\psi}_{mM}$ can be expressed as 
\begin{equation}
\label{gattononintegrato}
{\cal E}_{\psi_{mM}}=\sqrt{1-e^{-2\modul{\alpha}^2(1-\cos\varphi)}}.
\end{equation} 
In the range $\varphi\in[0,\pi]$ and regardless of $\modul{\alpha}$, %%@
Eq.~(\ref{gattononintegrato}) is a monotonously increasing function of the phase $\varphi$ %%@
saturating at ${\cal E}_{\psi_{mM}}(\modul{\alpha}\gg{1})=1$. The saturation value of this %%@
function depends on $\modul{\alpha}$ and, as a function of the interaction phase $\varphi$, is %%@
reached faster as $\modul{\alpha}$ is increased.
\begin{figure}[t]
{\hskip0.5cm}{\bf (a)}\hskip4cm{\bf (b)}
\psfig{figure=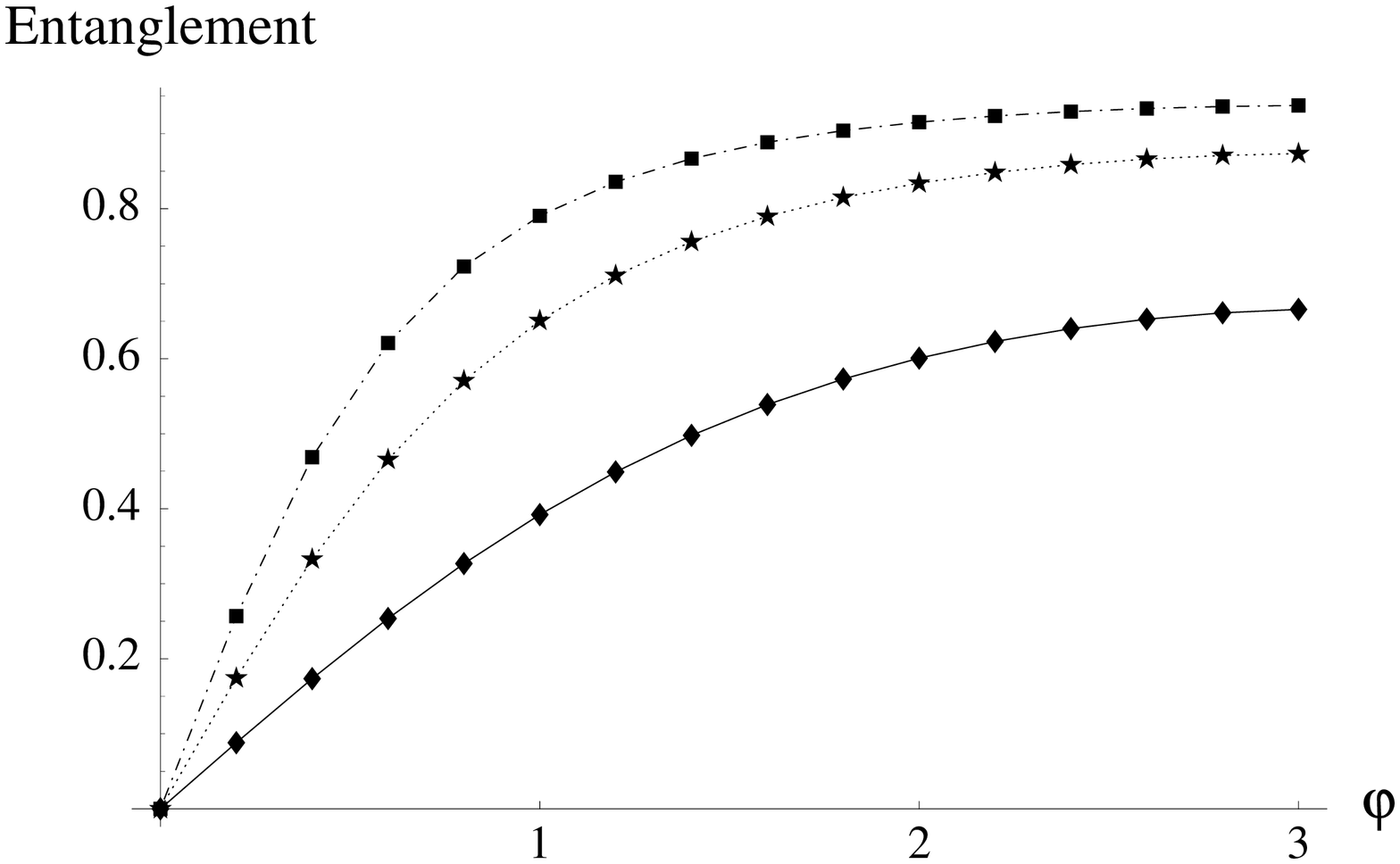,width=4.5cm,height=3.5cm}\psfig{figure=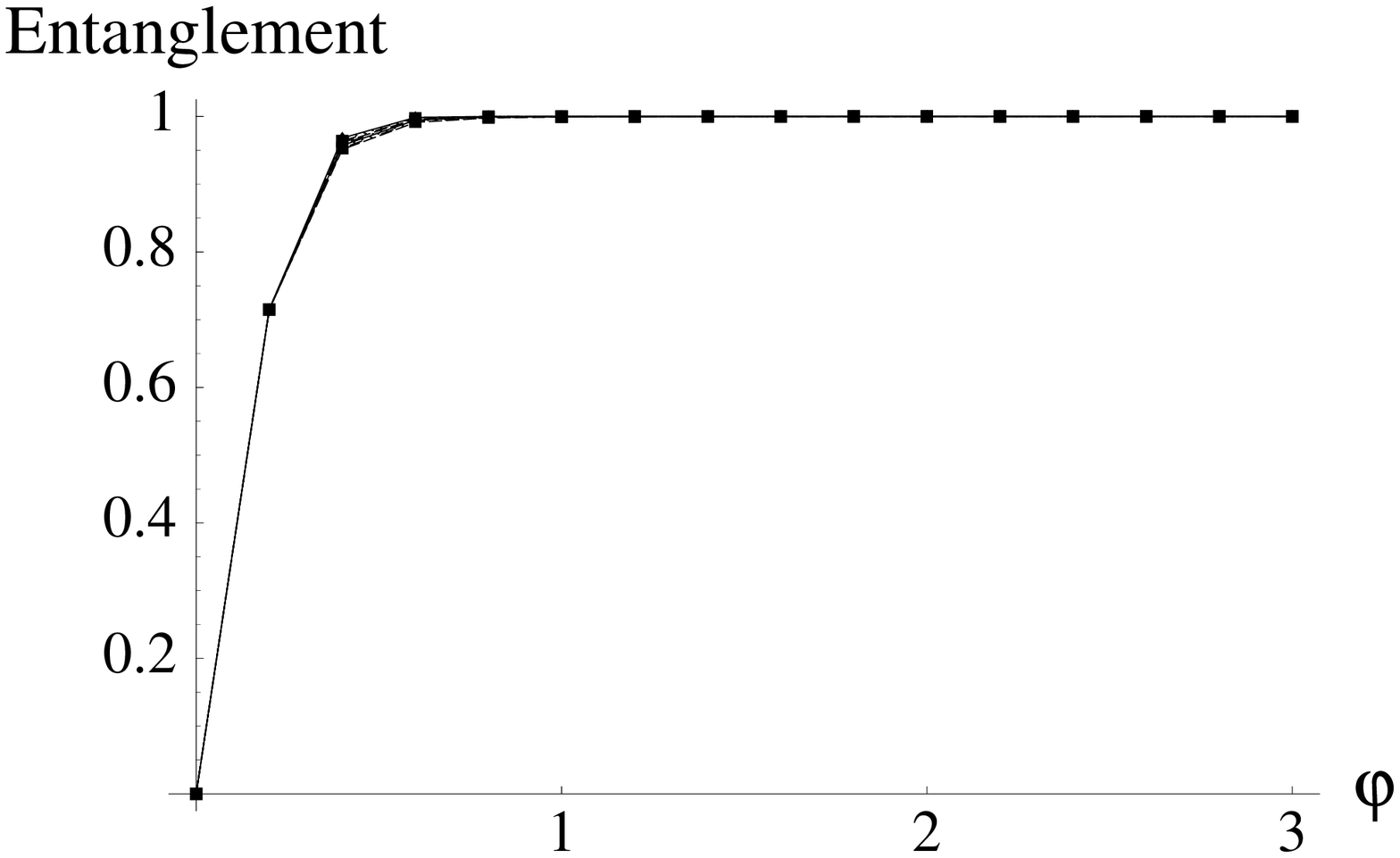,width=4.5cm,height=3.5cm}
\caption{Entanglement in the generalized cat-like state $\rho^{ent}_{mM}$ against the %%@
interaction phase $\varphi$, for different values of the variance $V$. In panel ${\bf (a)}$ we have considered the displacement $d=1$. From bottom to top, the curves show the behavior of the entanglement as $V$ is increased. We have considered $V=2$ ($\blacklozenge$), $V=5$ ($\bigstar$) and $V=10$ ($\blacksquare$). Panel $\bf{(b)}$ shows the results for $d=7$ %%@
showing that, for larger $d$, the curves relative to different $V$'s get so close to become indistinguishable. The %%@
saturation to one ebit of entanglement is nearly independent of the temperature of the thermal %%@
distribution.}
\label{gattointegrato}
\end{figure}
The Schmidt decomposition of $\ket{\psi}_{mM}$ is a useful tool in order to test in a simple %%@
way the entanglement present in the thermally weighted superposition $\rho^{ent}_{mM}$. By %%@
applying the NPT criterion, we find that ${\cal %%@
E}_{\rho^{ent}_{mM}}=2\int{d}^{2}\alpha{P}^{th}_{M}(V,d)\sqrt{\lambda_{-}\lambda_{+}}$ which, %%@
involving the double Gaussian integration of the square root of an exponential function is, in %%@
general, hard to compute analytically. Nevertheless, it has been possible to numerically %%@
sample the behavior of this function against the interaction phase $\varphi$, for different %%@
values of the variance $V$ and the displacement $d$. The results are shown in %%@
Figs.~\ref{gattointegrato} {\bf (a)} and {\bf (b)}, where $d=1$ and $d=7$ have been %%@
considered, respectively. Counterintuitively, the thermal weighting of the superposition does %%@
not smear out the entanglement in the bimodal state. Significant entanglement can be found for %%@
any value of the variance $V$ ({\it i.e.} at any value of the temperature considered). By %%@
looking at Fig. \ref{gattointegrato} {\bf (a)} we see that, as the variance $V$ increases, %%@
more entanglement is found in the state in Eq.~(\ref{eq3}). Moreover, as soon as a large %%@
displacement $d$ is considered, the differences between the results corresponding to %%@
distributions characterized by increasing variances becomes irrelevant. The saturation value %%@
of entanglement is reached regardless of the actual value of $V$ (see %%@
Fig.~\ref{gattointegrato} {\bf (b)}), quicker than the case of a small $d$. This matches the %%@
results found in terms of the negativity of the Wigner function associated with the state %%@
$\rho^{ent}_{mM}$ which, for moderate values of $V$, deepens its negative part for larger %%@
values of $d$. 

On the other hand, this effect represents a huge practical advantage in the production of the %%@
entangled state we are studying. Indeed, it is well-known that the currently achievable rates %%@
of non-linearity are not sufficiently large to guarantee an interaction phase %%@
$\varphi=\pi$~\cite{commentoEIT}. This would signify that, for a small $V$, the produceable %%@
entanglement would be very faint (see Fig.~\ref{gattointegrato} {\bf (a)}). However, it is %%@
sufficient to displace the thermal field more in order to achieve a significant improvement in %%@
the rate of entanglement generation. 
This is in line with the idea of using weak nonlinearities for various applications %%@
\cite{SNL,commentojacob}. Thus, our scheme seems to be experimentally accessible even in the %%@
more realistic situation of a small Kerr non-linearity.

We believe that this result is important in the context of entanglement in mesoscopic systems %%@
at finite temperatures~\cite{lavorivlatko3,sougato3} as we have shown a model where the %%@
quantum nature of a microscopic resource is enough in order to induce intrinsically quantum %%@
features in a macroscopic object (as our displaced thermal field). These features, which are %%@
temperature-resilient, can be highlighted by partitioning the CV Hilbert space into %%@
quadruplets at fixed $\alpha$ and averaging over the temperature-dependent probability %%@
distribution.

%%%%%%%%%%%%%%%%%%%%%%%%%%%%%%%%%%%%%%%%%%%%%%%%%%%%TRASFERIMENTO%%%%%%%%%%%%%%%%%%%%%%%%%%%%%%%@
%%%%%%%%%%%%%%%%%%%%%%%%
\section{Transfer of quantum properties to microscopic objects}
\label{trasferisco}

In this Section we address a relevant question related to the possibility of transferring the %%@
quantum correlations established in the generalized cat-like state $\rho^{ent}_{mM}$ to the %%@
initially separable bipartite system of two non-interacting qubits. This problem is worth %%@
addressing under many viewpoints. {\it In primis}, would the entanglement set in the %%@
generalized cat-like state be useful if a quantum channel has to be realized? In the context %%@
of distributed QIP, this is a relevant question as it has been shown that reliable channels, %%@
exhibiting genuine quantum features, are an irremissible resource for the performances of %%@
quantum computation~\cite{DQIP}. On the other hand, the interface between hetero-dimensional %%@
systems is {\it per se} a hot topic which has attracted a considerable attention, recently, %%@
especially focused onto the transfer of entanglement from a CV system to one which {\it lives} %%@
in a discrete Hilbert space~\cite{ioMEMS,noi2}. Finally, while the tomographic reconstruction %%@
of the properties of a CV system is a hard task to perform, its discrete-variable counterpart %%@
may be accomplished with much easier experimental protocols~\cite{sackett}. It would be thus %%@
desirable to design a protocol which allows one to infer the entanglement within a CV state %%@
without relying on its direct tomography. In a discrete-variable system this is possible %%@
through, for instance, entanglement witnesses detected with a minimal number of measurements %%@
\cite{toth}.

Such a protocol is provided by the research for the entanglement generated in a %%@
discrete-variable bipartite subsystem by the CV one via some {\it local} qubit-CV %%@
interaction. This gives a sufficient criterion for the entanglement between the field %%@
modes~\cite{ioMEMS}. Indeed, if the two CV modes are separable, there is no way that, through %%@
simple local interactions, quantum correlations could be established. This strategy has been %%@
already proven to be efficient in revealing the entanglement between two CV modes which have %%@
been fed into two spatially separated cavities and interacted with a pair of qubits. The %%@
question of there being any entanglement left between the cavity field modes after the %%@
interaction with the qubits has found a positive quantitative answer through the analysis %%@
of the {\it entangling power}. This is the strategy we would like to use here.

Of course, the quantitative results will depend on the model chosen for the local mode-qubit %%@
interaction. However, the implementation of an arbitrary model for light-matter interaction is %%@
not a trivial point. Specifically, it is a physical setup-dependent issue (meaning that %%@
certain physical setups enable the implementation of certain interaction models more %%@
straightforwardly than other). It is the physical system that one has in mind which dictates %%@
the most suitable form of the local interaction. Nevertheless, recently it has been shown that %%@
the standard Jaynes-Cummings (JC) model for resonant (as well as dispersive) qubit-boson %%@
interaction is common to many different physical realization of a quantum device and in a %%@
range of frequency from microwave to optical \cite{noi}. This dresses with physical %%@
significance the choice of the interaction %%@
$\hat{H}_{ij}=\Omega(\hat{a}_{i}\hat{\sigma}^{+}_{j}+h.c)$ [$(i,j)=(m,a)$ for the first field %%@
mode-qubit system and $(i,j)=(M,b)$ for the second] to govern the local dynamics of each %%@
qubit-field mode subsystem. Here, %%@
$\hat{\sigma}^{+}_{j}=\hat{\sigma}^{-\dagger}_{j}=\ket{1}_{j}\!\bra{0}$ is the raising %%@
operator of the $j$-th qubit (ordered logical basis $\{\ket{0},\ket{1}\}$) and $\Omega$ is a %%@
coupling strength. In the qubit computational basis, this interaction is the generator of the %%@
propagator
\begin{equation}
\label{JCrisolto}
\hat{U}_{ij}(\tau)=e^{-i\hat{H}_{ij}t}=
\begin{pmatrix}
\cos(\tau\sqrt{\hat{a}^{\dag}\hat{a}})&-i\hat{a}^{\dag}
\frac{\sin(\tau\sqrt{\hat{a}^{\dag}\hat{a}+1})}{\sqrt{\hat{a}^{\dag}\hat{a}+1}}\\
-i\hat{a}\frac{\sin(\tau\sqrt{\hat{a}^{\dag}\hat{a}})}{\sqrt{\hat{a}^{\dag}\hat{a}}}
&\cos(\tau\sqrt{\hat{a}^{\dag}\hat{a}+1})
\end{pmatrix}
\end{equation}
with $\tau=\Omega{t}$ being a rescaled interaction time. 
%Thus, the light-qubit state $\ket{0,n}$ evolves as
%$$\ket{0,n}_{ji}\stackrel{\hat{U}_{ij}(t)}{\rightarrow}\cos\theta_{n}\ket{0,n}
%_{ji}-i\sin\theta_{n}\ket{1,n-1}_{ji}.$$

%Again, a more powerful test for the entanglement in 
%$\ket{\psi(\alpha,\varphi)}=(1/\sqrt{2})\ket{0,\alpha}+\ket{1,e^{i\varphi}\alpha}$
%is provided by mapping the CV state onto the finite-dimensional Hilbert space of two qubits. 
It is convenient to proceed with the entangling power test by previously considering the %%@
initial separable state $\ket{\psi}_{mM}\otimes\ket{00}_{ab}$ and their joint dynamics %%@
$\hat{U}_{ma,Mb}=\hat{U}_{ma}\otimes\hat{U}_{Mb}$. 
%You can easily check that the global state, at the rescaled time $\tau$, reads
%\begin{equation}
%\begin{split}
%&\ket{\psi'(\alpha,\varphi,\tau)}=e^{-\frac{\modul{\alpha}^2}{2}}\sum^{\infty}_{n=0}
%\frac{\alpha^{n}}
%{\sqrt{n!}}\{[\ket{0,n}+e^{i\varphi{n}}\cos(\tau)\ket{1,n}]_{12}\cos(\tau\sqrt{n})
%\ket{0,0}_{ab}-i[\ket{0,n-1}\\
%&+e^{i\varphi{n}}\cos(\tau)\ket{1,n-1}]_{12}\sin(\tau\sqrt{n})\ket{01}
%_{ab}-ie^{i\varphi{n}}\sin\tau\cos(\tau\sqrt{n})\ket{0,n}_{12}\ket{10}
%_{ab}-e^{i\varphi{n}}\sin\tau\sin(\tau\sqrt{n})\ket{0,n-1}_{12}\ket{11}_{ab}\}.
%\end{split}
%\end{equation}
Tracing out the CV degrees of freedom, we are left with the two-qubit density matrix
\begin{equation}
\label{matricedensita}
\begin{aligned}
\rho_{\psi}(\tau)&=\mbox{Tr}_{mM}[\hat{U}_{ma,Mb}\ket{\psi}_{mM}
\!\bra{\psi}\otimes\ket{00}_{ab}\!\bra{00}\hat{U}^{-1}_{ma,Mb}]\\
&=e^{-{\modul{\alpha}^2}{}}\sum^{\infty}_{n=0}\frac{\modul{\alpha}^{2n}}{n!}
\begin{pmatrix}
A&iG_{1}&iG_{2}&-G_{3}\\
-iG^{}_{1}&B_{1}&F_{1}&iF_{2}\\
-iG^{}_{2}&F^{}_{1}&B_{2}&iH_{1}\\
-G^{}_{3}&-iF^{}_{2}&-iH^{}_{1}&B_{3}
\end{pmatrix}
\end{aligned}
\end{equation}
with the explicit expressions for the density matrix elements
\begin{equation}
\label{nonintegratedelements}
\begin{split}
&A=[1+{\mbox c}^{2}_1(\tau)]{\mbox c}^{2}_{n}(\tau),{\hskip0.2cm}G_{1}
=\frac{1+e^{-i\varphi}{\mbox c}^{2}_{1}(\tau)}{\sqrt{n+1}}\alpha^{*}
{\mbox c}_{n}(\tau){\mbox s}_{n}(\tau),\\
&G_{2}=e^{-i\varphi{n}}{\mbox s}_{0}(\tau)
{\mbox c}^{2}_{n}(\tau),\hskip0.2cm{G_{3}}=\frac{e^{-i\varphi(n+1)}
{\mbox s}_{0}(\tau)}{\sqrt{n+1}}\alpha^{*}{\mbox s}_{n}(\tau){\mbox c}_{n}(\tau),\\
&B_{1}=\frac{1+\cos^{2}\tau}{n+1}\modul{\alpha}^{2}{\mbox s}^{2}_{n}
(\tau),\hskip0.2cm{F_{1}}=\frac{e^{-i\varphi{n}}{\mbox s}_{0}(\tau)}
{\sqrt{n+1}}\alpha{\mbox c}_{n}(\tau){\mbox s}_{n}(\tau),\\
&{B_{2}}={{\mbox s}^{2}_{0}(\tau)}{\mbox c}^{2}_{n}(\tau),
\hskip0.2cm{F}_{2}=\frac{e^{-i\varphi(n+1)}{\mbox s}_{0}(\tau)}
{n+1}\modul{\alpha}^2{\mbox s}^{2}_{n}(\tau)\\
&B_{3}=\frac{{\mbox s}^{2}_{0}(\tau)}{n+1}\modul{\alpha}^2
{\mbox s}^{2}_{n}(\tau),\hskip0.2cm{H}_{1}=\frac{e^{-i\varphi}
{\mbox s}^{2}_{0}(\tau)}{\sqrt{n+1}}\alpha^{*}
{\mbox c}_{n}(\tau){\mbox s}_{n}(\tau).
\end{split}
\end{equation}
Here we have used the notation ${\mbox c}_{n}(\tau)=\cos(\tau\sqrt{n}),\,{\mbox %%@
s}_{n}(\tau)=\sin(\tau\sqrt{n+1})$. Eqs.~(\ref{nonintegratedelements}) have now to be %%@
integrated over the thermal weighting function in order to gather the reduce density matrix of %%@
the qubits after the interaction with $\rho^{ent}_{mM}$. By exchanging the summation in %%@
Eq.~(\ref{matricedensita}) and the Gaussian integrals and using the binomial formula, 
%\begin{equation}
%\modul{\alpha}^{2n}=(a^2+b^2)^n=\sum^{n}_{k=0}\frac{n!}{k!(n-k)!}a^{2k}b^{2(n-k)},
%\end{equation}
%where this time $\alpha=a+ib$ has been considered. In this way
evaluating the integral over $P^{th}(V,d)$ results in the introduction of the functions
\begin{equation}
\begin{split}
&\Theta^{j}_{n}(V)=\sum^{n+j}_{l=0}\left( {\begin{array}{*{20}c} n+j \\ l \\ \end{array}} %%@
\right)\left(\frac{V-1}{V+1}\right)^{l}\Gamma\left(l+\frac{1}{2}\right),\\
&\Lambda^{j}_{n}(V,d)=\sum^{2(n-l)+j}_{l=0}\left(
{\begin{array}{*{20}c} 2(n-l)+j \\ r \\ \end{array}} \right)\left(
\frac{2d}{V+1}\right)^{2(n-l)-r+j}\\
&\times\left(\frac{V-1}{V+1}\right)^{r/2}\frac{1}{2}(1+(-1)^r)
\Gamma\left(\frac{r+1}{2}\right),\hskip0.4cm(j=0,1,2)
\end{split}
\end{equation}
where $\left( {\begin{array}{*{20}c} n \\ l \\ \end{array}} \right)$ is the symbol for the %%@
binomial coefficient and $\Gamma(r)$ is the Gamma function of its argument $r$ but in a %%@
density matrix of the same form of Eq.~(\ref{matricedensita}).
\begin{figure}[t]
\psfig{figure=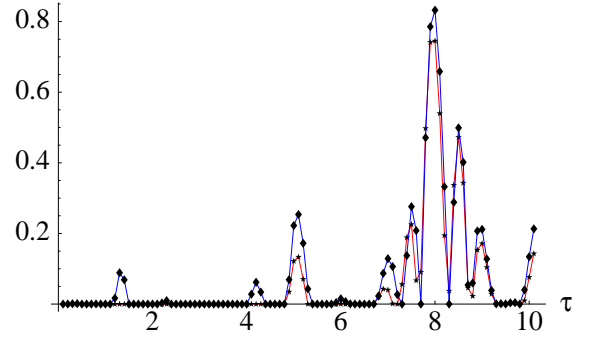,width=8.0cm,height=5cm}
\caption{Entangling power of the state $\int{d}^2{P}^{th}_{M}(V,d)\rho_{\psi}(\tau)$ for %%@
$V=10$ and $d=10$ ($\blacklozenge$, blue line) and for $V=10$ with $d=7$ ($\bigstar$, red %%@
line). In both the simulation, $\varphi=\pi$ has been assumed. The horizontal axis shows the %%@
rescaled interaction time $\tau=\Omega{t}$.}
\label{trasferito}
\end{figure}
The explicit eigenvalues of the integrated density matrix are not very informative as their %%@
form is quite complicated~\cite{commentodensita}. Nevertheless, it is possible to evaluate %%@
them numerically for a considerable range of $V$'s and $d$'s. Some striking results are %%@
presented in Fig.~\ref{trasferito} for $V=10$ (corresponding to an average photon number %%@
$\bar{n}=4.5$) and displacement $d=7$ ($\bigstar$) and $d=10$ ($\blacklozenge$). It is evident %%@
that the same effect shown in Fig.~\ref{gattointegrato} {\bf (b)}, regarding the increase in %%@
$d$, is achieved in the amount of entanglement transferred to the initially separable qubits. %%@
It is particularly interesting to stress the considerably large amount of transferred %%@
entanglement. In this situation a significantly entangled two-qubit channel can thus be %%@
constructed by exploiting the CV entangler represented by $\rho^{ent}_{mM}$.
\begin{figure}[t]
\psfig{figure=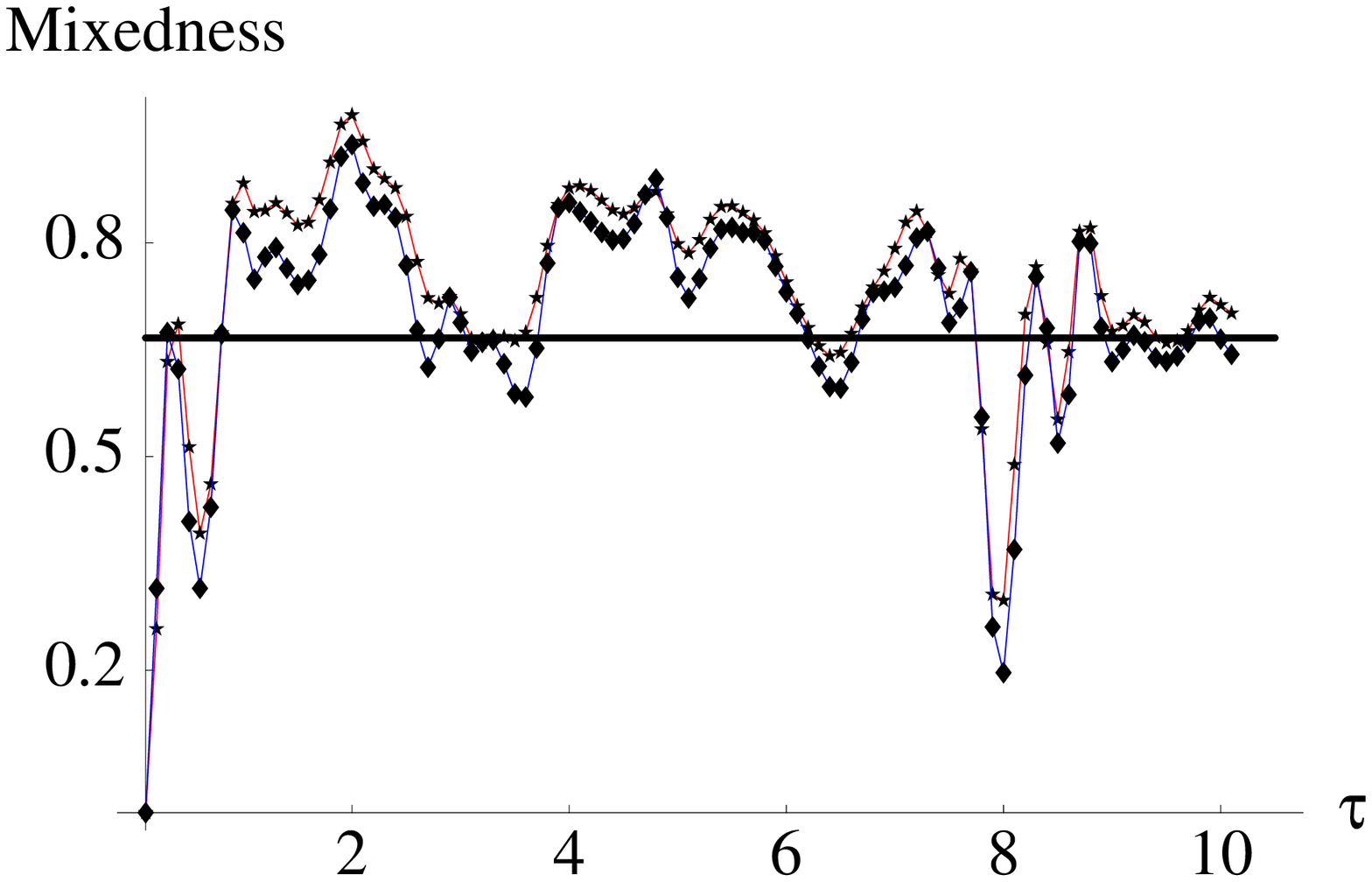,width=8.0cm,height=5cm}
\caption{Mixedness of the state $\int{d}^2{P}^{th}_{M}(V,d)\rho_{\psi}(\tau)$ against the %%@
rescaled interaction time $\tau$ for the same parameters as Fig.~\ref{trasferito}. Also shown %%@
the mixedness-threshold for a two-qubit state for quantum teleportation (thick straight line).}
\label{purezza}
\end{figure}
The usefulness of such a channel has to be quantified by including, in this analysis, the %%@
mixedness properties of the two-qubit state. Indeed, it is known that, for instance, a %%@
bipartite mixed state becomes useless for quantum teleportation whenever its linearized %%@
entropy $S_{l}=(4/3)\left[1-Tr\left(\rho^2_{12}(r,t)\right)\right]$ exceeds %%@
$2/{3}$~\cite{sougatovlatko}. $S_{l}$ is a good measure for mixedness which ranges from 0 (for %%@
pure states) to 1 (for maximally mixed ones). The calculation of the linearized entropies %%@
corresponding to the examples reported in Fig.~\ref{gattointegrato} and their comparison to %%@
the threshold for quantum teleportation is presented in Fig.~\ref{purezza}, showing the %%@
high-entanglement bumps which corresponds to a sufficiently pure state. The channel, in this case, 
can be faithfully used %%@
for quantum teleportation protocols. 

\begin{figure}[t]
\hskip1cm{\bf (a)}\hskip4cm{\bf (b)}
\psfig{figure=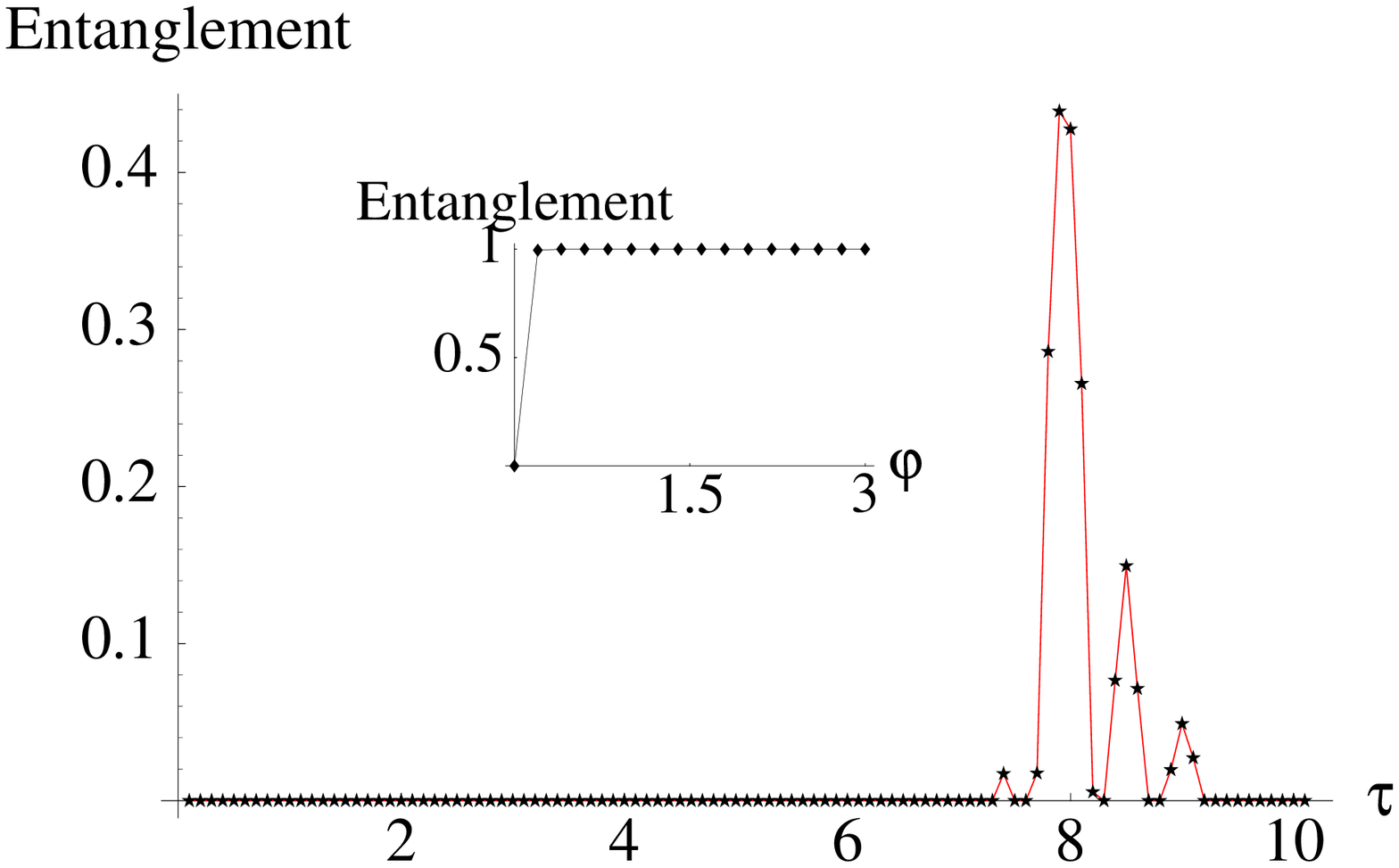,width=4.5cm,height=3.0cm}\psfig{figure=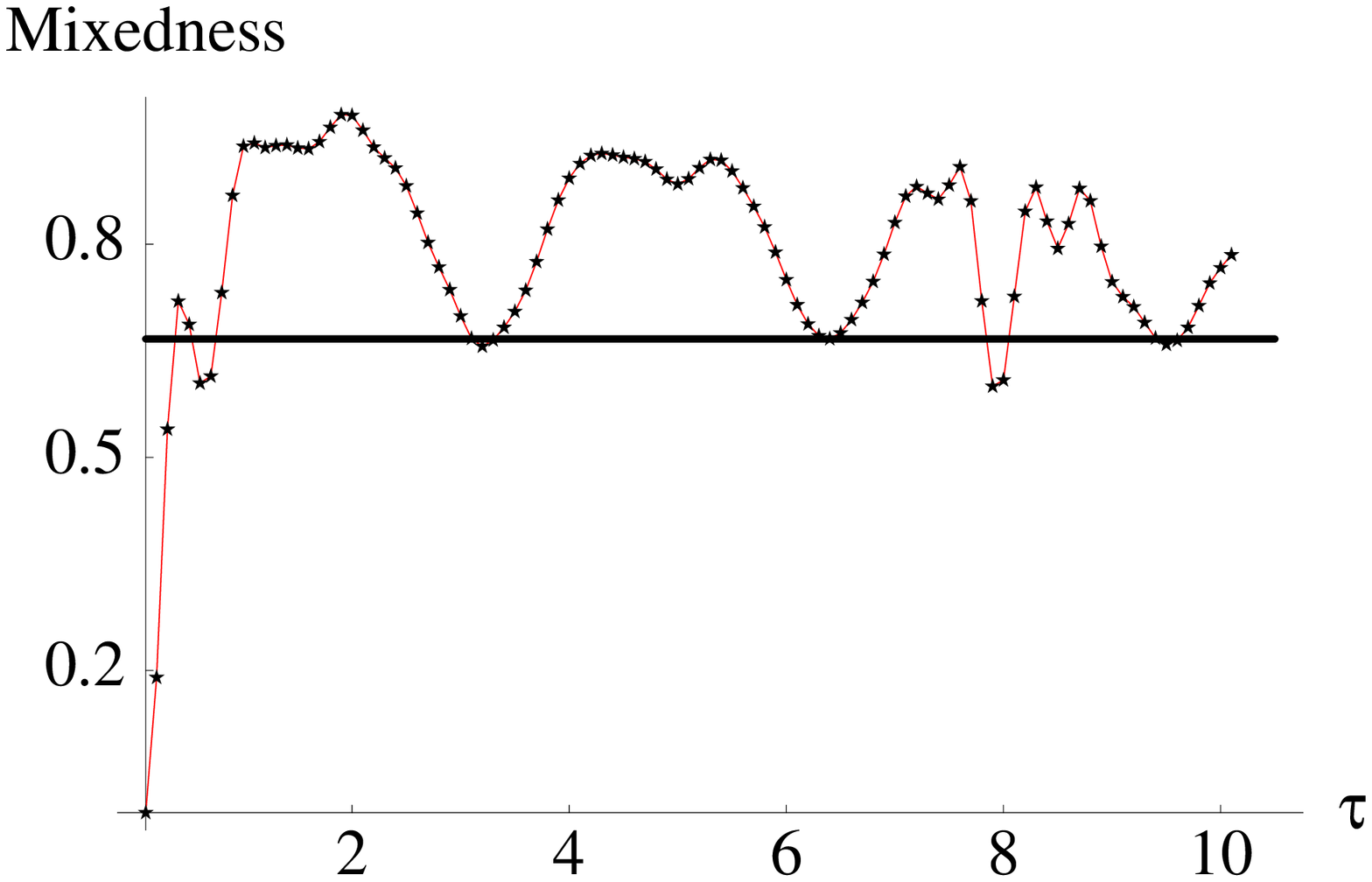,width=4.5cm,height=3.0cm}
\caption{{\bf (a)}: Entangling power of the state %%@
$\int{d}^2{P}^{th}_{M}(V,d)\rho_{\psi}(\tau)$ against $\tau$ for $V=100$ and $d=20$. The inset %%@
shows the entanglement, against $\varphi$, in the corresponding $\rho^{ent}_{mM}$. {\bf (b)}: %%@
Mixedness of the same state as a function of the rescaled interaction time. The %%@
mixedness-threshold for teleportation is also shown (straight line).}
\label{insieme}
\end{figure}

It is worth stressing the non-monotonic behavior of the entangling power against the %%@
entanglement initially present in the CV state. This is a well-known feature of this %%@
entanglement test~\cite{ioMEMS} and is a result of the interference of the Rabi flopping %%@
induced by the distribution of photons characterizing each state having a specific value of %%@
$V$ and $d$. As a specific instance of this peculiar non-linear relation between the amount of %%@
entanglement initially contained in the CV entangler and what is finally found in the %%@
two-qubit reduced state, we consider the case of $V=100$ with $d=20$. The analysis conducted %%@
in Section~\ref{gattogenerale} results in an entanglement which rapidly approaches $1$, as %%@
shown in the inset of Fig.~\ref{insieme} {\bf (a)}, a behavior quite consistent with the trend %%@
shown in Fig.~\ref{gattointegrato} {\bf (b)}. On the other hand, the entanglement transferred %%@
to the two-qubit state is never larger that $0.45$, as evidenced by Fig.~\ref{insieme} {\bf %%@
(a)}, which is smaller than the entanglement transferred to the qubits for $V=10$, $d=7$. We %%@
believe this result is still extremely significant as the high value of $V$ considered here %%@
(corresponding to $\bar{n}=49.5$ photons) shows that quite a considerable entangling power is %%@
in a high-temperature generalized cat-like state. Coming back to the example of teleportation %%@
we have addressed, the corresponding two-qubit channel is still useful as its mixedness, at %%@
$\tau=8$, can be well-below the threshold (see Fig.~\ref{insieme} {\bf (b)}).

%%%%%%%%%%%%%%%%%%%%%%SETUP%%%%%%%%%%%%%%%%%%%%%%%%%%%%

\section{Proposal for the experimental verification}
\label{experiment}

In the previous Section we have theoretically addressed the question of how to infer the %%@
entanglement between the subsystems of a generalized cat-like state. 
On a practical side, however, if our attention is restricted to the case of travelling-wave %%@
fields, some problems have to be faced.

In the first place, we have already stressed the difficulties related to the achievement of %%@
large-enough rates of non-linearity. In Section~\ref{gattogenerale} we have outlined a %%@
strategy to highlight the quantum correlations in generalized cat-like states generated with %%@
smaller interaction phases $\varphi$'s. Nevertheless, the realization of even small %%@
$\varphi$'s would pass, for instance, through the use of long optical fibers where the effect %%@
of dephasing channel is still an unknown issue~\cite{commentojacob}.

\begin{figure}[t]
%\centerline{\includegraphics[width=0.4\textwidth]{Presentation1.bmp}}
\psfig{figure=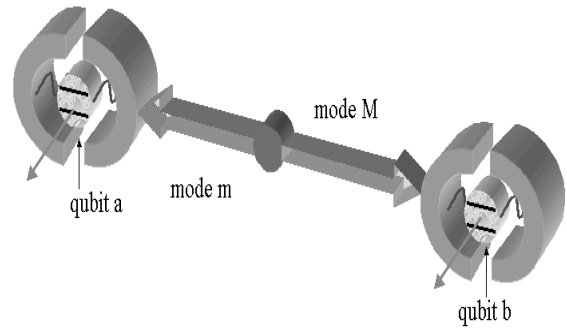,width=7.5cm,height=4.8cm}
\caption{Scheme of principle of the experiment to be realized in order to test the entangling %%@
power as described in Section~\ref{trasferisco}. Modes $m$ and $M$ are assumed to be %%@
addressing two cavities through which two independent qubit pass. The cavity-qubit %%@
interaction changes the refraction index of the cavity, allowing for the feeding by the %%@
external fields.}
\label{esperimento}
\end{figure}

In the second place, the realization of the entangling power test would require a demanding %%@
(even if foreseeable) experimental setup. In Fig.~\ref{esperimento} we sketch the scheme 
of the idea put forward in Section~\ref{trasferisco}. The generalized cat-like state %%@
being embodied by two travelling field modes following the general scheme of %%@
Fig.~\ref{schema}, would feed two optical cavities, each crossed by a two-level atom (or %%@
containing an integrated quantum dot, as an alternative). The passage of the atoms would set %%@
each cavity field mode in resonance with the external driving mode, which will penetrate the %%@
cavity and interact with the two-level system through the dynamics described in %%@
Eq.~(\ref{JCrisolto}). Each step required by this experimental setup has been independently %%@
demonstrated (see ref.~\cite{ioMEMS} and references within for a detailed discussion) and we %%@
will not comment further on them. On the other hand, the main point of this Section would be %%@
the introduction of a simplified scheme which is still able to highlight the important %%@
features of our study with a much more realistic physical setup. 

In our idea, the most striking difference with %%@
respect to the previous configuration is the presence of a single cavity and no travelling %%@
light field. Indeed, it is well-known~\cite{effettiva} (see also Paternostro {\it et al.} %%@
\cite{mqip}) that the dispersive interaction between a single two-level system and a cavity %%@
mode would lead to an effective Hamiltonian reading
\begin{equation}
\label{hamiltonianaeffettiva}
\hat{H}^{eff}_{mM}=\chi{\hat{M}^{\dag}\hat{M}}\ket{1}_{a}\!\bra{1},
\end{equation}
where the microscopic subsystem $m$ has been identified with the qubit $a$ whose spectrum has %%@
been rescaled so that its ground state has zero energy. We have indicated %%@
$\chi=\Omega^2/\delta$ with $\delta\gg{\Omega}$ the large atom-field detuning. This model can %%@
be achieved by assuming a qubit $1$-cavity mode JC interaction in the presence of a static %%@
electric field. The induced Stark shift on the atomic levels creates a detuning $\delta$ such %%@
that Eq.~(\ref{hamiltonianaeffettiva}) holds. 
%\begin{figure}[ht]
%\psfig{figure=schema2.eps,width=7.3cm,height=5.5cm}
%\caption{Scheme for the experimental verification of the analysis presented in this paper. The %%@
%scheme reproduces the single-qubit transfer protocol we address in the body of the paper. In %%@
%this case, the generalized cat-like state is given by the state of qubit $1$ and the cavity %%@
%mode field. Qubit $2$ is ancillary and is used just in order to test the entangling power in a %%@
%less demanding way.}
%\label{esperimento2}
%\end{figure}
The propagator generated by this Hamiltonian, starting from the state %%@
$\ket{+}_{a}\otimes\rho^{th}_{M}(V,d)=(1/\sqrt %%@
2)(\ket{0}+\ket{1})_{a}\otimes\rho^{th}_{M}(V,d)$, creates a generalized cat-like state %%@
between a thermal displaced state of the cavity field mode and a two-level atom. The %%@
displacement of the thermal field can be effectively performed via the modification to the %%@
cavity refractive index induced by the presence of the two-level atom. This acts as an %%@
effective non-linear intra-cavity medium which shifts the resonance frequency of the cavity %%@
and can inject an external coherent state of its amplitude $\beta$. The effective displacement %%@
would result in $d=\beta\Delta{t}$ with $\Delta{t}$ the time-of-flight for atom $a$ crossing the %%@
cavity~\cite{davidovich}.

We are thus considering a hybrid cat-like state where the CV and the qubit parts have clearly %%@
distinguished physical embodiments. As soon as qubit $a$ leaves the cavity, a second qubit %%@
$b$, identical to the first, crosses the cavity in absence of the Stark electric field %%@
({\it i.e.} the interaction model is the standard JC one). By interacting with the cavity %%@
field, this process leads to an entangled state of subsystems $a,\,M$ and $b$. Tracing out the %%@
cavity field, we are left with a reduced two-qubit state whose entanglement represents, again, %%@
a sufficient condition for the entanglement in the generalized cat-like state. We are thus %%@
describing a unilateral entanglement-transfer process for a modified entangling power %%@
test~\cite{commentoswapping}.  

Obviously, the quantitative analysis reported in Section \ref{trasferisco} are not applicable to %%@
the present modified protocol and the effectiveness of the entanglement transfer process %%@
should be re-tested. However, this is a straightforward process which may be derived directly %%@
from what is presented in Eqs.~(\ref{matricedensita}) and~(\ref{nonintegratedelements}). We %%@
find that the structure of the reduced density matrix %%@
$\rho_{\psi,ab}={\mbox{Tr}}_{M}(\hat{U}_{Mb}\hat{U}^{eff}_{mM}\rho^{th}_{M}\otimes\ket{+}_{a}\!\bra{+}\otimes\ket{g}_{b}\!\bra{g}\hat{U}^{eff\dag}_{mM}\hat{U}^{\dag}_{Mb})$ is similar to %%@
Eq.~(\ref{matricedensita}).
% with the following matrix elements
%\begin{equation}
%\begin{aligned}
%A_{int}&=\frac{e^{-\frac{2d^2}{V+1}}}{\pi(V+1)}\sum^{\infty}_{n=0}
%\frac{{\mbox c}^{2}_{n}(\tau)}{n!}{\Theta}^{0}_{n}(V)\Lambda^{0}_{n}(V,d),\\
%G_{1int}&=i\frac{e^{-\frac{2d^2}{V+1}}}{\pi(V+1)}\sum^{\infty}_{n=0}
%\frac{{\mbox c}_{n}(\tau){\mbox s}_{n}(\tau)}{n!\sqrt{n+1}}
%\Theta^{0}_{n}(V)\Lambda^{1}_{n}(V,d),\\
%G_{2int}&=\frac{e^{-\frac{2d^2}{V+1}}}{\pi(V+1)}\sum^{\infty}_{n=0}
%\frac{{\mbox c}^2_{n}(\tau)e^{-in\varphi}}{n!}\Theta^{0}_{n}(V)\Lambda^{0}_{n}(V,d),\\
%G_{3int}&=i\frac{e^{-\frac{2d^2}{V+1}}}{\pi(V+1)}
%\sum^{\infty}_{n=0}\frac{{\mbox c}_{n}(\tau)
%{\mbox c}_{n}(\tau)e^{-i(n+1)\varphi}}{n!\sqrt{n+1}}\Theta^{0}_{n}(V)\Lambda^{1}_{n}(V,d),\\
%B_{1int}&=\frac{e^{-\frac{2d^2}{V+1}}}{\pi(V+1)}\sum^{\infty}_{n=0}
%\frac{{\mbox s}^2_{n}(\tau)}{(n+1)!}\Theta^{1}_{n}(V)\Lambda^{2}_{n}(V,d),\\
%F_{1int}&=-i\frac{e^{-\frac{2d^2}{V+1}}}{\pi(V+1)}
%\sum^{\infty}_{n=0}\frac{{\mbox c}_{n}(\tau)
%{\mbox s}_{n}(\tau)e^{-in\varphi}}{n!\sqrt{n+1}}\Theta^{0}_{n}(V)\Lambda^{1}_{n}(V,d),\\
%F_{2int}&=\frac{e^{-\frac{2d^2}{V+1}}}{\pi(V+1)}\sum^{\infty}_{n=0}\frac{
%{\mbox s}^2_{n}(\tau)e^{-i(n+1)\varphi}}{(n+1)!}\Theta^{1}_{n}(V)
%\Lambda^{2}_{n}(TV,d),\\
%H_{1int}&=i\frac{e^{-\frac{2d^2}{V+1}}}{\pi(V+1)}\sum^{\infty}_{n=0}
%\frac{{\mbox c}_{n}(\tau)
%{\mbox s}_{n}(\tau)e^{-i\varphi}}{n!\sqrt{n+1}}\Theta^{0}_{n}(V)
%\Lambda^{1}_{n}(V,d),
%\end{aligned}
%\end{equation}
\begin{figure}[ht]
\hskip1cm{\bf (a)}\hskip4cm{\bf (b)}
\psfig{figure=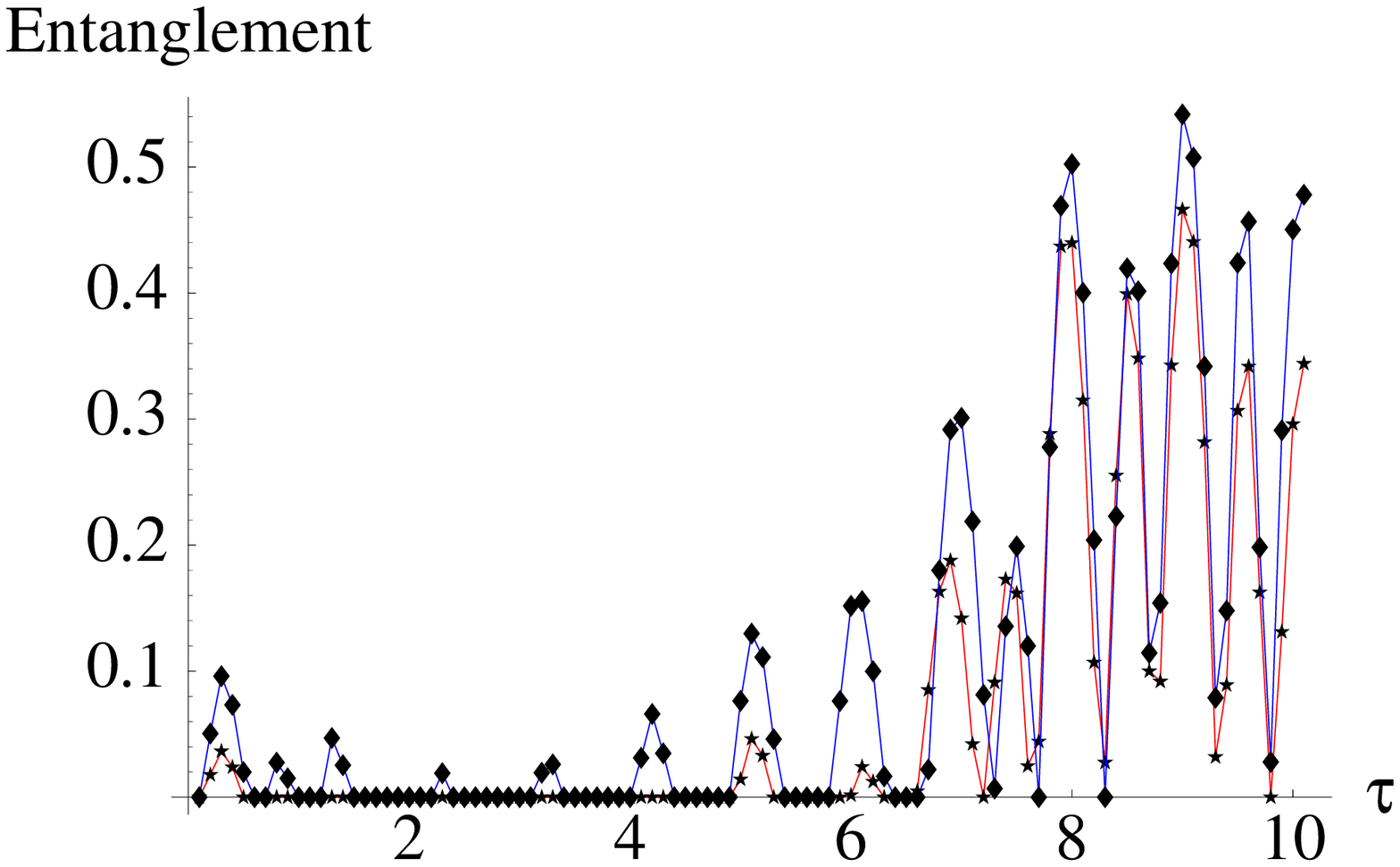,width=4.5cm,height=3.0cm}\psfig{figure=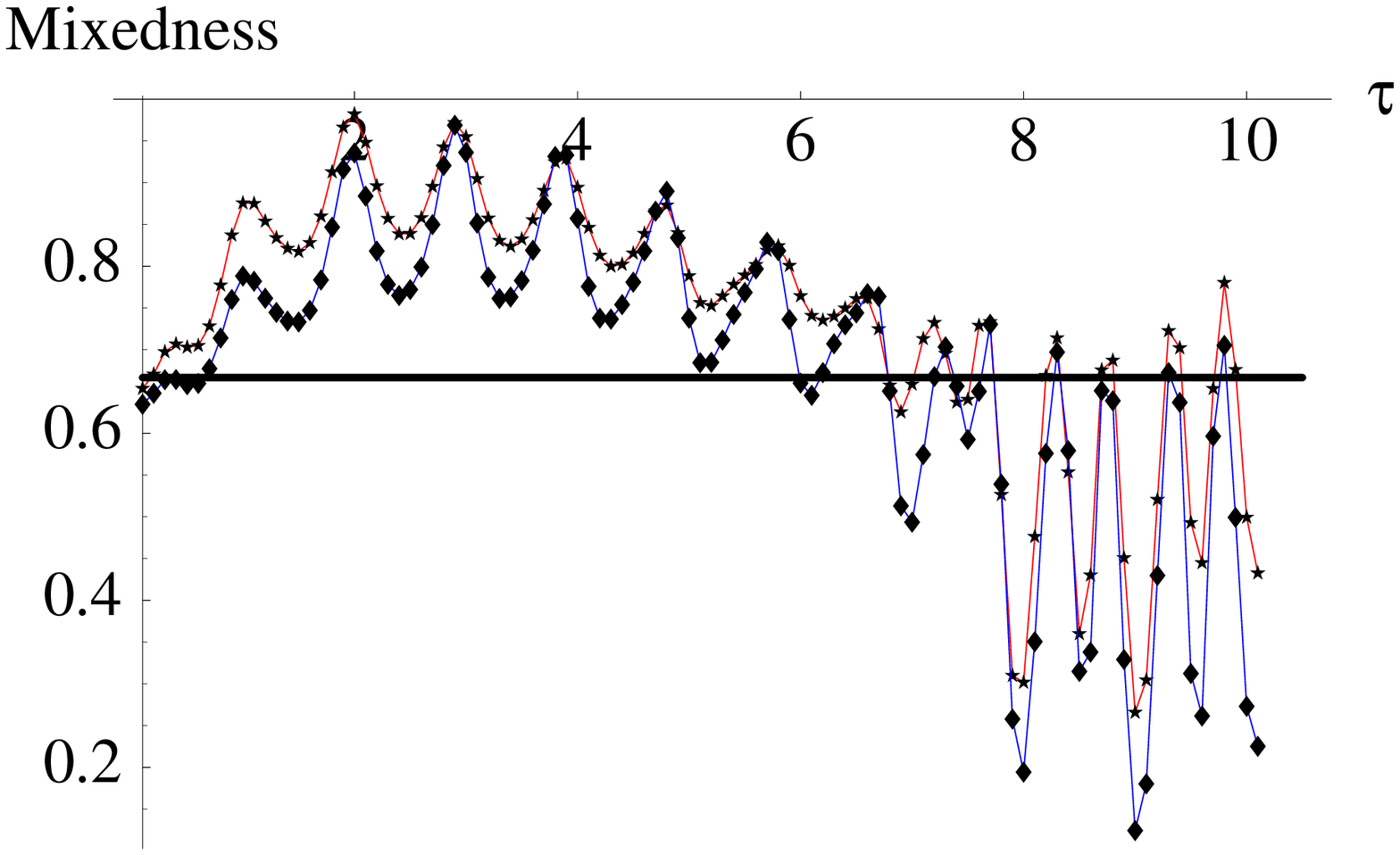,width=4.5cm,height=3.0cm}
\caption{{\bf (a)}: Entangling power of the state %%@
$\int{d}^2{P}^{th}_{M}(V,d)\rho_{\psi}(\tau)$ against $\tau$ for $V=10$ and $d=10$ %%@
($\blacklozenge$) and $V=10$ with $d=7$ ($\bigstar$). {\bf (b)}: Mixedness of the same state %%@
as a function of the rescaled interaction time. The mixedness-threshold for teleportation is %%@
also shown (straight line).}
\label{insieme2}
\end{figure}
%with $B_{2int}=A_{int},\,B_{3int}=B_{1int}$. 
The calculation of the negativity of partial %%@
transposition gives us, once again, useful information about the entanglement within this %%@
two-qubit state, transferred after the $M-b$ interaction.
The results are shown in Fig.~\ref{insieme2} {\bf (a)} for the same parameters as in %%@
Fig.~\ref{trasferito}. At the same time, in Fig.~\ref{insieme2} {\bf (b)} we reprise our %%@
teleportation example and present the analysis in terms of linearized entropy. It is evident %%@
that a good degree of entanglement can still be found (despite the high peaks of %%@
Fig.~\ref{trasferito} have disappeared), with a quite low degree of mixedness of the %%@
associated quantum channel also for this modified entangling test protocol. Obviously, this %%@
represents a huge advantage in terms of experimental feasibility as all the ingredients for %%@
the entangling power test are within the current state of the art.   

%%%%%%%%%%%%%%%%%%%%%%%%%REMARKS%%%%%%%%%%%%%%%%%%%%%%%%%%%%%%%%%%%%
\section{Remarks}
\label{conclusioni}

In investigating the entangled properties of mesoscopic systems at finite temperature, in the %%@
quantum optics domain, we often face some difficulties related to a certain lack of analytical %%@
tools. This has strongly limited the insight we could gain about the specific behavior, in %%@
terms of quantum correlations, of particular classes of states (especially non-Gaussian %%@
states).

These difficulties have forced the researchers to look for ways to bypass the problem and %%@
infer the quantumness of a state. It appears that such the alternative tests are based on %%@
the restriction of the analysis from infinite to finite dimensional Hilbert spaces. This can be %%@
done by effective (formal) projection of a CV state onto bipartite Hilbert sectors (as done in %%@
refs.~\cite{lavorivlatko3,sougato2}) as well as looking at the amount of non-classical %%@
correlations that a given state under investigation can transfer to mutually non-interacting %%@
qubits~\cite{ioMEMS}. In this paper we have performed some interesting steps along these %%@
directions, investigating the entanglement properties of a class of recently introduced %%@
(highly non-Gaussian) generalized cat-like states~\cite{jacob2}.
Our approach has been twofold. In the first place, we have calculated the entanglement of the %%@
cat-like state by partitioning the Hilbert space of the combined system into manifolds spanned %%@
by the states of an effective bipartite two-level system. The entanglement of the whole state, %%@
then, is the result of a thermal average of the quantum correlations within each manifold. The %%@
qualitative features highlighted by this approach have been confirmed, in the second place, by %%@
testing the entangling power of the generalized cat-like state with respect to two initially %%@
separable qubits. This second approach is particularly useful under a practical as well as a %%@
theoretical viewpoint. It has revealed that powerful experimental criteria exist, for the %%@
entanglement investigation, beyond the CV criteria for inseparability~\cite{simon}. Our study %%@
is, thus, a particularly illuminating case of this second, more operative scenario. 

As an additional example of the effectiveness of the entangling power test, here we briefly %%@
assess the problem of the entanglement in the class of states generated by the entire protocol %%@
described by the setup in Fig.~\ref{schema} (dashed-box included). The detection of mode $m$ %%@
is performed onto the basis $\{\ket{+},\ket{-}\}$ (with %%@
$\ket{-}=(1/\sqrt{2})(\ket{0}-\ket{1})$). Conditioned on finding $\ket{\pm}_{m}$, the action %%@
of the beam-splitter which mixes mode $M$ to vacuum gives rise to the state (we assume %%@
$\varphi=\pi$)
\begin{equation}
\label{doposplitter}
\begin{split}
\rho^\pm&={\cal N}^{\pm}\int{d}^2\alpha{P}^{th}(V,d)\left\{
\pro{\sigma,-\sigma}{\sigma,-\sigma}+\pro{\sigma,-\sigma}{-\sigma,\sigma}\right.\\
&\left.\pm\pro{-\sigma,\sigma}{\sigma,-\sigma}\pm\pro{-\sigma,\sigma}
{-\sigma,\sigma}\right\},
\end{split}
\end{equation}
where $\sigma=\alpha/\sqrt{2}$ and ${\cal N}^{\pm}$ are proper normalization factors. In order %%@
to fix the ideas, let us focus on the case of $\ket{+}_m$ being found with $d=0$, for which %%@
the expression of the variance matrix is particularly straightforward, reading
\begin{equation}
\gamma^{+}=
\begin{pmatrix}
\frac{V^2+1}{2V}&0&-\frac{(V-1)^2}{2V}&0\\
0&\frac{V^2+1}{2V}&0&-\frac{(V-1)^2}{2V}\\
-\frac{(V-1)^2}{2V}&0&\frac{V^2+1}{2V}&0\\
0&-\frac{(V-1)^2}{2V}&0&\frac{V^2+1}{2V}
\end{pmatrix}.
\end{equation}
It is easy to check that Simon's separability condition is not violated by %%@
Eq.~(\ref{doposplitter}) so that no firm statement about the entanglement properties of this %%@
state can be made. However, following the lines depicted in ref.~\cite{jacob2}, it is %%@
straightforward to show that Eq.~(\ref{doposplitter}) violates a Bell-CHSH %%@
inequality~\cite{bell} regardless of $V$ and $d$, thus revealing an inherent quantum entanglement %%@
of the state. This result is confirmed by the application of the entangling power test which %%@
shows that entanglement (as large as $\simeq{0.6}$ for V around $50$ and $d=5$) can be %%@
transferred to two mutually non-interacting qubits using the symmetric scheme of %%@
Section~\ref{trasferisco}. This second example completes and complements our investigation which, we believe, represents %%@
an interesting example of the possibilities of revealing quantum non-locality in non-trivial %%@
situations of mesoscopic superpositions at high temperature. 
\\
%\vskip2cm
\acknowledgments

We acknowledge useful discussions with Prof. P. L. Knight and Dr. J. Fiurasek and communication %%@
with A. Ferreira. This work has been supported by The Leverhulme Trust (ECF/40157), the %%@
Australian Research Council, the UK EPSRC and the Korea Research Foundation (2003-070-C00024).
\\


\begin{thebibliography}{99}
\bibitem{commentodensita} Some advantages, in looking for an analytical expression for the %%@
eigenvalues come from the fact that, for $\varphi=\pi$, $G_{1}=-H_1$ and $G_2\simeq{-}F_2$, %%@
which considerably simplify the expression of $\int{d}^2{P}^{th}_{M}(V,d)\rho_{\psi}(\tau)$.
\bibitem{commentoswapping} The process may also be seen as a variant of the %%@
entanglement-swapping protocol.
\end{thebibliography}
\end{document}